\documentclass[11pt,a4paper,fleqn]{article}
\catcode`\@=11
\@addtoreset{equation}{section}
\def\theequation{\arabic{section}.\arabic{equation}}
\def\appendix{\renewcommand{\thesection}{\Alph{section}}\setcounter{section}{0}
              \renewcommand{\theequation}
            {\mbox{\Alph{section}.\arabic{equation}}}\setcounter{equation}{0}}

\def\maketitle{\thispagestyle{empty}\setcounter{page}0\newpage
                \renewcommand{\thefootnote}{\arabic{footnote}}
                  \setcounter{footnote}0}
\renewcommand{\thanks}[1]{\renewcommand{\thefootnote}{\fnsymbol{footnote}}
               \footnote{#1}\renewcommand{\thefootnote}{\arabic{footnote}}}

\renewcommand{\title}[1]{\begin{center}\Large\bf #1\end{center}\rm\par\bigskip}

\renewcommand{\author}[1]{\begin{center}\Large #1\end{center}}
\newcommand{\address}[1]{\begin{center}\large #1\end{center}}

\def\babs{\hrule\par\begin{description}\item{Abstract: }\it}
\def\eabs{\par\end{description}\hrule\par\medskip\rm}
\renewcommand{\date}[1]{\par\bigskip\par\sl\hfill #1\par\medskip\par\rm}

\def\segue{\qquad\Longrightarrow\qquad} 
\def\hs{\qquad}               
\def\nn{\nonumber}            
\def\beq{\begin{eqnarray}}    
\def\be{\begin{eqnarray}}
\def\eeq{\end{eqnarray}}      
\def\ee{\end{eqnarrayn}}
\def\ap{\left.}               
\def\at{\left(}               
\def\aq{\left[}               
\def\ag{\left\{}              
\def\cp{\right.}              
\def\ct{\right)}              
\def\cq{\right]}              
\def\cg{\right\}}             

\def\R{{\hbox{{\rm I}\kern-.2em\hbox{\rm R}}}}   
\def\H{{\hbox{{\rm I}\kern-.2em\hbox{\rm H}}}}   
\def\N{{\hbox{{\rm I}\kern-.2em\hbox{\rm N}}}}   
\def\C{{\ \hbox{{\rm I}\kern-.6em\hbox{\bf C}}}} 
\def\Z{{\hbox{{\rm Z}\kern-.4em\hbox{\rm Z}}}}   
\def\ii{\infty}                                  
\renewcommand{\Re}{\mathop{\rm Re}\nolimits}       
\renewcommand{\Im}{\mathop{\rm Im}\nolimits}       
\def\dir{/\kern-.7em D\,}                          
\def\lap{\Delta\,}                                 
\def\al{\alpha}
\def\ga{\gamma}
\def\de{\delta}
\def\ep{\varepsilon}
\def\ze{\zeta}

\def\la{\lambda}

\def\si{\sigma}

\def\Ga{\Gamma}

\def\be{\begin{equation}}
\def\ee{\end{equation}}
\def\bea{\begin{eqnarray}}
\def\eea{\end{eqnarray}}

\def\nn{\nonumber}
\def\e{{\rm e}}

\renewcommand{\title}[1]{\begin{center}\Large\bf #1\end{center}\rm\par\bigskip}
\renewcommand{\author}[1]{\begin{center}\Large #1\end{center}}

\begin{document}

\title{One-loop $f(R)$ gravity in de Sitter universe}
\author{Guido Cognola$\,^{(a)}$\thanks{cognola@science.unitn.it},
Emilio Elizalde$\,^{(b)}$\thanks{elizalde@ieec.fcr.es},
Shin'ichi Nojiri$\,^{(c)}$\thanks{nojiri@nda.ac.jp, snojiri@yukawa.kyoto-u.ac.jp},\\
Sergei D.~Odintsov$\,^{(b,d)}$\thanks{odintsov@ieec.fcr.es, also at TSPU,Tomsk}
and
Sergio Zerbini$\,^{(a)}$\thanks{zerbini@science.unitn.it}
}
\address{
$^{(a)}$ Dipartimento di Fisica, Universit\`a di Trento \\
and Istituto Nazionale di Fisica Nucleare \\
Gruppo Collegato di Trento, Italia\\
\medskip
$^{(b)}$ Consejo Superior de Investigaciones Cient\'{\i}ficas
(ICE/CSIC) \, and \\  Institut d'Estudis Espacials de Catalunya 
(IEEC) \\
Campus UAB, Fac Ciencies, Torre C5-Par-2a pl \\ E-08193 Bellaterra
(Barcelona) Spain\\
\medskip
$^{(c)}$ Department of Applied Physics,
National Defence Academy, \\
Hashirimizu Yokosuka 239-8686, Japan \\
\medskip
$^{(d)}$ ICREA, Barcelona, Spain \, and \\  
Institut d'Estudis Espacials de Catalunya (IEEC) \\
Campus UAB, Fac Ciencies, Torre C5-Par-2a pl \\ E-08193 Bellaterra
(Barcelona) Spain\\
}

\begin{abstract}
Motivated by the dark energy issue, the one-loop
quantization approach for a family of relativistic cosmological
theories is discussed in some detail. Specifically,
general $f(R)$ gravity at the one-loop level in a de Sitter
universe is investigated, extending a similar program developed 
for the case of pure Einstein gravity. Using generalized zeta 
regularization, the one-loop effective action is explicitly
obtained off-shell, what allows to study in detail the possibility of 
(de)stabilization of the de Sitter background by quantum effects. 
The one-loop effective action maybe useful also for the study 
of constant curvature black hole nucleation rate and it provides 
the plausible way of resolving the cosmological constant problem.

\end{abstract}

\maketitle

\section{Introduction}

Recent astrophysical data indicate that our universe is currently in a
 phase of accelerated expansion. One possible explanation for this fact
is to postulate that gravity is being nowadays modified by some terms
which grow when curvature decreases. This could be, for instance, an
inverse curvature term \cite{turner} which might have its origin at
a very fundamental level, as string/M-theory\cite{nojiri0} or the
presence of higher dimensions\cite{zhuk}, which seem able to explain
such accelerated expansion. Gravity modified with 
inverse curvature terms is known to contain some instabilities \cite{ins}
and cannot pass some solar system tests, but further modifications of the
same which include higher derivative, curvature squared
terms make it again viable \cite {NO2}. (The Palatini formulation may also improve
the situation, for recent discussion, see \cite{palatini} and refs. therein).

Having in mind possible applications of modified gravity for the
late time universe, the following question appears. If it so happens
that Einstein gravity is only an approximate theory, looking at the
early universe should this (effective) quantum gravity  be different
from the Einsteinian one? A widely discussed possibility in this
direction is quantum $R^2$ gravity (for a review, see \cite{buch}).
However, other modifications are welcome as well, because they
sometimes produce extra terms which may help to realize the early
time inflation. This is supported by the possibility of accelerated
expansion with simple modified gravity. Thus, we will study here
general $f(R)$ gravity at the one-loop level in a de Sitter
universe. A similar program for the case of  pure Einstein gravity
(recall that it is also multiplicatively non-renormalizable) has
been initiated in refs. \cite{perry, duff80,frad} (see also
\cite{ds1}). Using generalized zeta-functions regularization (see, for
instance \cite{eli94,byts96}), one can get the one-loop effective
action and then study the possibility of stabilization of the
de Sitter background by quantum effects. Moreover, such approach
hints also to a possible way of resolving the cosmological constant
problem \cite{frad}. Hence, the study of one-loop $f(R)$ gravity is a
natural step to be undertaken for the completion of such a program,
keeping always in mind, however, that a consistent quantum gravity
theory is not available yet. But in any case, one should 
also not forget that, from our present knowledge, current gravity, 
which might indeed deviate from Einstein's one, ought to have 
its origin in the Planck era and come from a more fundamental 
quantum gravity/string/M-Theory  approach.

Let us briefly review  the classical modified gravity theory which
depends only on scalar curvature: \be \label{XXX7} I={1 \over
\kappa^2}\int d^4 x \sqrt{-g} f(R)\ . \ee Here $\kappa^2=16 \pi G$ and $f(R)$ 
is, in principle, an arbitrary function. By introducing the auxiliary fields
$A$ and $B$, one may rewrite the action (\ref{XXX7}) as \be
\label{X1} I={1 \over \kappa^2}\int d^4 x \sqrt{-g}
\left[B\left(R-A\right) + f(A)\right]\ . \ee Using the equation of
motion and deleting $B$ one gets to the Jordan frame action. Using
the conformal transformation $g_{\mu\nu}\to \e^\sigma g_{\mu\nu}$
with $\sigma = -\ln f'(A)$, we obtain the Einstein frame action
(scalar-tensor gravity), as follows: \bea \label{XXX11} && I_E= {1
\over \kappa^2}\int d^4 x \sqrt{-g} \left\{ R - {3 \over 2}
\left[{f''(A) \over f'(A)}\right]^2
g^{\rho\sigma}\partial_\rho A \partial_\sigma A - {A \over f'(A)} + {f(A) \over f'(A)^2}\right\} \nn\\
&& \ \ \ \ \ ={1 \over \kappa^2}\int d^4 x \sqrt{-g} \left[ R - {3
\over 2}g^{\rho\sigma}
\partial_\rho \sigma \partial_\sigma \sigma - V(\sigma)\right]\ , \\
\label{XXX11b} && V(\sigma)= \e^\sigma g\left(\e^{-\sigma}\right) -
\e^{2\sigma} f\left(g\left(\e^{-\sigma} \right)\right) =  {A \over
f'(A)} - {f(A) \over f'(A)^2}\ . \eea Note that two such classical
theories, in these frames, are mathematically equivalent. It is
known that they are not equivalent however at the quantum level
(off-shell), due to the use of different parametrizations. Even at
the classical level, the physics they describe seems to be
different. For instance, in the Einstein frame matter does not seem
to freely fall along geodesics, what is a well established fact.

As an interesting and specific example of the general setting above,
the following action corresponding to gravity modified at large
distances may be considered \cite{turner,capo93} 
\be \label{BRX1} I={1
\over \kappa^2}\int d^4 x \sqrt{-g} \left(R - \mu R^{-n}\right)\ .
\ee Here $\mu$ is  (an extremely small) coupling constant and $n$ is
some number, assumed to be $n>-1$. The function $f(A)$ and the
scalar field $\sigma$ are \be \label{XXX13} f(A)=A -\mu A^{-n}\
,\quad \sigma=-\ln \left(1 + n\mu A^{-n-1}\right). \ee The
Friedman-Robertson-Walker (FRW) universe metric in the Einstein
frame is chosen as \be \label{X3} ds_E^2= -dt_E^2 + a_E^2(t_E)
\sum_{i=1}^3\left(dx^i\right)^2\ . \ee If the curvature is small,
the solution of  equation of motion is \cite{capo93}
\be \label{X4}  a_E \sim
t_E^\frac{3(n+1)^2}{(n+2)^2} \ ,\quad \sigma = \frac{n+1}{n+2} \ln
\frac{t_E}{t_{E_0}}\ , \quad t^2_{E_0}\equiv
\frac{\frac{(n+1)^2}{(n+2)^2}\left(1 -
\frac{3(n+1)^2}{4(n+2)^2}\right)} {2\left(1 +
\frac{1}{n}\right)\left(n \mu \right)^{\frac{1}{n+1}}}\ . \ee The
FRW metric in the Jordan frame is \be \label{X5} ds^2= -dt^2 +
a^2(t) \sum_{i=1}^3\left(dx^i\right)^2\ , \ee where the variables in
the Einstein frame and in the physical Jordan frame are related with
each other by \be \label{X6} t= \int \e^{\frac{\sigma}{2}}dt_E\ ,
\quad a=\e^{\frac{\sigma}{2}}a_E\ , \ee which gives $t\sim t_E^{1
\over n+2}$ and \be \label{XXX14b} a\sim t^{(n+1)(2n+1) \over n+2}\
,\quad w=-{6n^2 + 7n - 1 \over 3(n+1)(2n+1)}\ . \ee The first
important consequence of the above Eq.~(\ref{XXX14b}) is that there
is the possibility of accelerated expansion for some choices of $n$
(a kind of effective quintessence). In fact, if
 $n>\frac{-1+\sqrt{3}}{2}$ or
$-1<n<-\frac{1}{2}$, it follows that $w<-\frac{1}{3}$ and $\frac{d^2
a}{dt^2}>0$. This is the reason why such a theory \cite{turner} and
some  modifications thereof \cite{NO2,GRG,wang,ricci,fara} have been
indeed  widely considered as candidates for gravitational dark
energy models. The black holes and wormholes in such models have
been also discussed \cite{brevik, cognola}.

When $-1<n<-\frac{1}{2}$, one arrives at $w<-1$, i.e.
 the universe is shrinking. If we replace the
direction of time by changing $t$ by $-t$, the universe is expanding
but $t$ should be considered to be negative so that the scale factor
$a$ ought to be real. Then there appears a singularity at $t=0$,
where the scale factor $a$ diverges as $a \sim \left(  - t
\right)^\frac{2}{3(w+1)}$. One may shift the origin of time by
further changing $-t$ with $t_s - t$. Hence, in the present
universe, $t$ should be less than $t_s$ and a singularity is seen to
appear at $t=t_s$ (for a discussion of this point, see \cite{NO2}):
\be \label{X7} a \sim \left(t_s  - t \right)^\frac{2}{3(w+1)}\ . \ee
That is the sort of  Big Rip singularity. We should note that in the
Einstein frame (\ref{X3}), the solution (\ref{X4}) gives $w$ in the
Einstein frame as \be \label{X8} w_E=-1 + \frac{2(n+2)^2}{9(n+1)^2}
> -1\ . \ee Therefore, in the Einstein frame, there is no singularity
of the Big Rip type. In general, for the scalar field $\varphi$ with
potential $U(\varphi)$ and canonical kinetic term, the energy
density $\rho$ and the pressure $p$ are given by \be \label{X9} \rho
= \frac{1}{2}{\dot\varphi}^2 + U(\varphi)\ ,\quad p =
\frac{1}{2}{\dot\varphi}^2 - U(\varphi)\ . \ee Therefore $w$ is
given by \be \label{10} w=\frac{p}{\rho}= -1 +
\frac{{\dot\varphi}^2}{\frac{1}{2}{\dot\varphi}^2 +
 U(\varphi)} \ ,
\ee which is bigger than $-1$ if $U(\varphi)> -
\frac{1}{2}{\dot\varphi}^2$. Therefore in order that
 the phantom with $w<-1$ is realized, one needs a non-canonical kinetic
 term or a negative potential. For the action (\ref{XXX11}), the sign of
 the kinetic term for the scalar field $\sigma$ is the same
as that in the canonical action. Therefore, in order to obtain a
phantom in the Einstein frame, at least, we need $V(\sigma)<0$ or
$Af'(A)<f(A)$. For the case (\ref{BRX1}), if $\mu<0$, $V$ can be
negative. From (\ref{XXX13}), however, if the curvature $R=A$ is
small, $\sigma$ becomes imaginary if $n>0$, what indicates that the
curvature cannot become so small. This is quite general. Let us
assume that for a not  very small curvature, $f(R)$ could be given
by the Einstein action, $f(R)\sim R$; then, $f'(R)\sim 1>0$. On the
other hand, if we also assume that when the curvature is small,
$f(R)$ behaves as $f(R)\sim \mu  R^{-n}$, where $\mu <0$ and $n>0$,
then $f'(R)<0$ for small $R$. Hence, $f'(R)$ should vanish for
finite $R$, which tells us that $\sigma$ diverges since $\sigma=-\ln
f'(A)$. Therefore $R$ cannot become so small. Even in the case when
$0>n>-1$, Eq.~(\ref{X4}) tells us that $t_{E_0}$ and therefore
$\sigma$ becomes imaginary. Then a negative $\mu$ could be
forbidden. At least the region where we have effective Einstein
gravity does not seem to be continuously connected with the region
where curvature is small for the case when a negative potential from
modified gravity follows. Thus, classical modified gravity mainly 
supports the accelerated expansion.

The paper is organized as follows. In the next section the classical 
dynamics of $f(R)$ gravity in de Sitter space is considered.
Section three is devoted to the calculation of the one-loop effective action in $f(R)$ gravity in de Sitter space. The explicit off-shell and on-shell 
effective action is found. The study of quantum-corrected de Sitter geometry 
for specific model of modified gravity is done in section four.
It turns out that quantum gravity corrections shift the radius of de Sitter space trying to destabilize it.
The calculation of the entropy for de Sitter space and constant curvature black holes is done in section five. Some remarks about black hole nucleation rate are done. The Discussion section gives some summary and outlook.
In the Appendix A the black hole solutions with constant curvature
 are explicitly given for $f(R)$ gravity. Appendix B is devoted 
to the details of the calculation of functional determinants for
 scalars, vectors and 
tensors in de Sitter space.

\section{Modified gravity models}

Recall that the  general relativistic theory  we are interested in is
 described by the action
\be \nonumber 
I=\frac{1}{16 \pi G}\int d^4x \sqrt{-g} f(R) \,,
\label{genR} \ee 
$f$ being a function of scalar curvature only.

We have many possible simple choices for the Lagrangian density
$f(R)$. Here we consider the following three examples: 
\be
f(R)=R+pR^2-2\Lambda\,, \label{q} \ee 
that is Einstein's gravity with
quadratic corrections; \be f(R)= R-\frac{\mu_1}{R} \,,\hs\hs\mu_1>0\,,
\label{invR} \ee the model proposed in Ref.~\cite{turner} and its
trivial generalization \be f(R)=R-\frac{\mu_1}{R} -\mu_2\,.
\label{invR12} \ee Here we shall be  interested in models which
admit solutions with constant 4-dimensional curvature $R=R_0$, an
example being the one of de Sitter. The general equations of motion
for the model described by Eq.~(\ref{genR}) are (see, for example,
\cite{barrow83}) \be f'(R)R_{\mu\nu}-\frac{1}{2} f(R)g_{\mu\nu}+\at
\nabla_\mu \nabla_\nu-g_{\mu\nu} \nabla^2 \ct f'(R)=0\,,
\label{geneqR} \ee $f'(R)$ being the derivative of $f(R)$ with
respect to $R$. If we now require the existence of solutions with
constant scalar curvature $R=R_0$, we arrive at \be
f'(R_0)R_{\mu\nu}=\frac{f(R_0)}{2}g_{\mu\nu}\,. \label{geneqRc} \ee
Taking the trace, we have the condition \cite{barrow83} \be
2f(R_0)=R_0\,f'(R_0) \label{B} \ee and this means that the solutions
are Einstein's spaces, namely they have to satisfy the equation \be
R_{\mu\nu}=\frac{f(R_0)}{2 f'(R_0)}g_{\mu\nu}=
\frac{R_0}{4}g_{\mu\nu}\,, \label{E} \ee $R_0$ being a solution of
Eq.~(\ref{B}). This gives rise to the effective cosmological
constant: \be \Lambda_{eff}=\frac{f(R_0)}{2
f'(R_0)}=\frac{R_0}{4}\,. \label{l} \ee For the  model defined by
Eqs. (2.2) and (\ref{invR}) we have, respectively, \be
R_0=4\Lambda\,,\hs \hs \Lambda_{eff}=\Lambda\,, \ee \be
R_0^2=3\mu_1\,,\hs \hs \Lambda_{eff}=\pm\sqrt{3\mu_1}\,. \ee It is clear
that this class of constant curvature solutions contains the
4-dimensional black hole solutions in the presence of a non
vanishing cosmological constant, like the Schwarzschild-(anti)de
Sitter solutions and all the topological solutions associated with a
negative $\Lambda_{eff}$. In particular, with $\Lambda_{eff}>0$,
there exist the de Sitter and Nariai solutions, while for
$\Lambda_{eff}<0$ there exists the anti de Sitter solution. For the
sake of completeness, in Appendix \ref{solution} we shall review all
of them.

\section{Quantum field fluctuations around the maximally symmetric instantons}

In this Section we will discuss the one-loop quantization of the
model on the a maximally symmetric  space. Of course this should be
considered only an effective approach (see, for instance
\cite{buch}). To start, we recall that the action describing a
generalized Euclidean gravitational theory is \beq
I_E[g]=-\frac{1}{16\pi G}\int\:d^4x\,\sqrt{g}\,f(R)\,, \eeq where
the generic function $f(R)$ satisfies --on shell-- the condition
\beq f'(R_0)=\frac{2f(R_0)}{R_0}\,, \label{AAA1} \eeq which ensures
(as we have seen) the existence of constant curvature solutions. In
particular, we are interested in instantons with constant scalar
curvature $R_0$, being also maximally symmetric spaces, namely
having covariant conserved curvature tensors. An  example is the
$S(4)$ de Sitter instanton. For maximally symmetric space, we have
\beq R^{(0)}_{ijrs}=\frac{R_0}{12}\at g^{(0)}_{ir}g^{(0)}_{js}-
g^{(0)}_{is}g^{(0)}_{jr}\ct \:, \hs
R^{(0)}_{ij}=\frac{R_0}{4}\,g^{(0)}_{ij}\,, \hs
R=R_0=\frac{12}{a^2}\,. \label{AAA2} \eeq Another symmetric
background is $H(4)$, associated with a negative cosmological
constant. For the $S(2) \times S(2)$ instanton, the consideration we
are going to extract are not valid, because it is not a maximally
symmetric space and Eq.~(\ref{AAA2}), which we shall use several
times from now on, does not hold true. In that case, one should make
use of the techniques described in refs. \cite{ginspar,wipf}. To
start with, let us consider small fluctuations around the maximally
symmetric instanton \beq g_{ij}=g^{(0)}_{ij}+h_{ij}\:,\hs
g^{ij}=g^{(0)ij}-h^{ij}+h^{ik}h^j_k+{\cal O}(h^3)\:,\hs
h=g^{(0)ij}h_{ij}\:. \eeq As usual, indices are lowered and raised
by the metric $g^{(0)}_{ij}$. Up to second order in $h_{ij}$, one
has \beq
\frac{\sqrt{g}}{\sqrt{g^{(0)}}}=1+\frac12h+\frac18h^2-\frac14h_{ij}h^{ij}
+{\cal O}(h^3)  \eeq and \beq
 R &\sim& R_0-\frac{R_0}{4}\,h+\nabla_i\nabla_jh^{ij}-\lap h
\nn \\ && +\frac{R_0}{4}\,h^{jk}h_{jk} -\frac14\,\nabla_ih\nabla^ih
-\frac14\,\nabla_kh_{ij} \nabla^kh^{ij} +\nabla_ih^i_k\nabla_jh^{jk}
-\frac12\,\nabla_jh_{ik}\nabla^ih^{jk} \:, \eeq where $\nabla_k$
represents the covariant derivative in the unperturbed metric
$g^{(0)}_{ij}$.

By performing a Taylor expansion of $f(R)$ around $R_0$, again up to
second order in $h_{ij}$, we get 
\beq S_E[g]=-\frac{1}{16\pi
G}\int\:d^4x\,\sqrt{-g^{(0)}}\: \aq f_0+\frac14(2f_0-R_0f'_0)\,h
+{\cal L}_2\,\cq\,, \label{AAA3} 
\eeq 
where,  up to total derivatives,
\begin{eqnarray}
{\cal L}_2&=&\frac12\:f''_0\:h_{ij}\,\nabla_i\nabla_j\nabla_r\nabla_s\,h_{rs}
\nn\\&&\hs
+\frac{1}{12}\,h_{ij}\aq3f'_0\lap-3f_0+R_0f'_0\cq\:h_{ij}
-\frac12\,f'_0\,h_{ij}\nabla_i\nabla_k\,h_{jk}
\nn\\&&\hs\hs
-\frac14\,h\aq +4f''_0\lap-2f'_0+R_0f''_0\cq\nabla_i\nabla_j\,h_{ij}
\nn\\&&
+\frac{1}{96}\,h\aq48f''_0\lap^2-24\at f'_0-R_0f''_0\ct\lap
+12f_0-8R_0f'_0+3R_0^2f''_0\cq\,h\:.
\end{eqnarray}
For the sake of simplicity,  we have used the notation
$f_0=f(R_0)$, $f'_0=f'(R_0)$ and $f''_0=f''(R_0)$, and in
what follows we will set $X\equiv \frac14(2f_0-R_0f'_0)$
(deviation from the on-shell condition).

It is convenient to carry out the standard expansion of the tensor
field $h_{ij}$ in irreducible components \cite{frad}, namely \beq
h_{ij}=\hat
h_{ij}+\nabla_i\xi_j+\nabla_j\xi_i+\nabla_i\nabla_j\sigma
+\frac14\,g_{ij}(h-\lap_0\sigma)\:, \label{tt}\eeq where $\si$ is
the scalar component, while $\xi_i$ and $\hat h_{ij}$ are the vector
and tensor components with the properties \beq
\nabla_i\xi_i=0\:,\hs\hs \nabla_i\hat h_{ij}=0\:,\hs\hs \hat
h_{ii}=0\:. \label{AAA4} \eeq In terms of the irreducible components
of the $h_{ij}$ field, the Lagrangian density, again disregarding
total derivatives, becomes
\begin{eqnarray}
{\cal L}_2&=&
\frac{1}{12}\,\hat h^{ij}\,(3f'_0\lap_2-3f_0+R_0f'_0\,)\,\hat h_{ij}
\nn\\&&\hs
+\frac1{16}\,(2f_0-R_0f'_0)\:\xi^i\,(4\lap_1+R_0)\,\xi_i
\nn\\&&
+\frac{1}{32}\:h\aq 9f''_0\,\lap_0^2
   -3(f'_0-2R_0f''_0\,)\,\lap
     +2f_0-2R_0f'_0+R_0^2f''_0\,
\cq\,h
\nn\\&&
+\frac{1}{32}\:\si\,\aq 9f''_0\lap_0^4 -3(f'_0-2R_0f''_0\,)\,\lap_0^3\cp
\nn\\&&\hs\hs\ap
     -(6f_0-2R_0f'_0-R_0^2f''_0\,)\,\lap_0^2
      -R_0(2f_0-R_0f'_0\,)\,\lap_0
\cq\,\si
\nn\\&&
\frac{1}{16}\:h\aq -9f''_0\,\lap_0^3
     +3(f'_0-2R_0f''_0\,)\,\lap_0^2
      +R_0(f'_0-R_0f''_0\,)\,\lap_0
\cq\,\si\:,
\end{eqnarray}
where $\lap_0$,  $\lap_1$ and $\lap_2$ are the Laplace-Beltrami operators
acting on scalars, traceless-transverse vector and
traceless-transverse tensor fields  respectively.
The latter expression is valid off-shell. In obtaining such expression,
due to the huge number of terms appearing in the computation, we have used
a program for symbolic tensor  manipulations.

As it is well known, invariance under diffeormorphisms renders the
operator in the $(h,\si)$ sector not invertible. One needs a gauge
fixing term and a corresponding ghost compensating term. We consider
the class of gauge condition, parametrized by the real parameter
$\rho$, \beq \chi_k=\nabla_j h_{jk}-\frac{1+\rho}4\,\nabla_k\,h\:,
\nn\eeq the harmonic or de Donder one corresponding to the choice
$\rho=1$. As gauge fixing,  we choose the quite general term
\cite{buch} \beq {\cal
L}_{gf}=\frac12\,\chi^i\,G_{ij}\,\chi^j\,,\hs\hs
G_{ij}=\al\,g_{ij}+\beta\,g_{ij}\lap\,, \label{AAA5} \eeq where the
term proportional to $\al$ is the one normally used in Einstein's
gravity. The corresponding ghost Lagrangian reads \cite{buch} \beq
{\cal L}_{gh}= B^i\,G_{ik}\frac{\de\,\chi^k}{\de\,\ep^j}C^j\,,
\label{AAA6} \eeq where $C_k$ and $B_k$ are the ghost and anti-ghost
vector fields, respectively, while $\de\,\chi^k$ is the variation of
the gauge condition due to an infinitesimal gauge transformation of
the field. It reads \beq
\de\,h_{ij}=\nabla_i\ep_j+\nabla_j\ep_i\segue
\frac{\de\,\chi^i}{\de\,\ep^j}=g_{ij}\,\lap+R_{ij}+\frac{1-\rho}{2}\,\nabla_i\nabla_j\,.
\label{AAA7} \eeq Neglecting total derivatives, one has \beq {\cal
L}_{gh}=B^i\,\at\al\,H_{ij}+\beta\,\lap\,H_{ij}\ct\,C^j\,,
\label{AAA8} \eeq where we have set \beq
H_{ij}=g_{ij}\at\lap+\frac{R_0}{4}\ct+\frac{1-\rho}{2}\,\nabla_i\nabla_j\,.
\label{AAA9} \eeq In irreducible components, one obtains
\begin{eqnarray}
{\cal L}_{gf} &=&\frac{\al}2\aq\xi^k\,\at\lap_1+\frac{R_0}4\ct^2\,\xi_k
    +\frac{3\rho}{8}\,h\,\at\lap_0+\frac{R_0}3\ct\,\lap_0\,\si
\cp\nn\\&&\hs\ap
    -\frac{\rho^2}{16}\,h\,\lap_0\,h
-\frac{9}{16}\,\si\,\at\lap_0+\frac{R_0}3\ct^2\,\lap_0\,\si
\cq
\nn\\&&
+\frac{\beta}2\aq\xi^k\,\at\lap_1+\frac{R_0}4\ct^2\,\lap_1\xi_k
    +\frac{3\rho}{8}\,h\,\at\lap_0+\frac{R}4\ct\at\lap_0+\frac{R}3\ct\,
\lap_0 \si
\cp\nn\\&&\hs\ap
    -\frac{\rho^2}{16}\,h\,\at\lap_0+\frac{R_0}4\ct\,\lap_0 h
    -\frac{9}{16}\,\si\,\at\lap_0+\frac{R_0}4\ct\at\lap_0+
\frac{R_0}3\ct^2\,\lap_0 \si \cq\,, \label{AAA10} \eeq \beq {\cal
L}_{gh} &=&\alpha\ag\hat B^i\at\lap_1+\frac{R_0}{4}\ct\hat C^j
+\frac{\rho-3}{2}\,b\,\at\lap_0-\frac{R_0}{\rho-3}\ct\,\lap_0 c\cg
\nn\\&&\hs +\beta\ag\hat
B^i\,\at\lap_1+\frac{R_0}{4}\ct\,\lap_1\,\hat C^j \cp
\nn \\
  &&\hs\hs\ap +\frac{\rho-3}{2}\,b\,\at\lap_0+\frac{R_0}{4}\ct
   \at\lap_0-\frac{R_0}{\rho-3}\ct\,\lap_0 c\cg\,,
\eeq
where ghost irreducible components are defined by
\beq
C_k&=&\hat C_k+\nabla_k c\,,\hs\hs \nabla_k\hat C^k=0\,,
\nn\\
B_k&=&\hat B_k+\nabla_k b\,,\hs\hs \nabla_k\hat B^k=0\,.
\label{AAA11} \eeq In order to compute the one-loop contributions to
the effective action one has to consider the path integral for the
bilinear part \beq {\cal L}= {\cal L}_2+\,{\cal L}_{gf}+{\cal
L}_{gh} \label{AAA12} \eeq of the total Lagrangian and take into
account the Jacobian due to the change of variables with respect to
the original ones. In this way, one gets  \cite{frad,buch} \beq
Z^{(1)}&=&\at\det G_{ij}\ct^{-1/2}\,\int\,D[h_{ij}]D[C_k]D[B^k]\:
\exp\,\at -\int\,d^4x\,\sqrt{g}\,{\cal L}\ct
\nn\\
&=&\at\det G_{ij}\ct^{-1/2}\,\det J_1^{-1}\,\det J_2^{1/2}\,
\nn\\&&\times \int\,D[h]D[\hat h_{ij}]D[\xi^j]D[\si] D[\hat
C_k]D[\hat B^k]D[c]D[b]\:\exp\, \at-\int\,d^4x\,\sqrt{g}\,{\cal
L}\ct\,, \eeq where $J_1$  and $J_2$ are the Jacobians due to the
change of variables in the ghost and tensor sectors respectively
\cite{frad}. They read \beq J_1=\lap_0\,,\hs\hs
J_2=\at\lap_1+\frac{R_0}{4}\ct\at\lap_0+\frac{R_0}{3}\ct\,\lap_0\,,
\label{AAA13} \eeq and the determinant of the operator  $G_{ij}$,
acting on vectors, can be written as \beq \det G_{ij}=\mbox{const}\
\det\at\lap_1+\frac{\al}{\beta}\ct\,\det\at\lap_0+\frac{R_0}4+\frac{\al}{\beta}\ct\,,
\label{AAA14} \eeq while it is trivial in the case $\beta=0$.

Now, a straightforward computation leads to the following off-shell
one-loop contribution to the ``partition function'' \beq
e^{-\Ga^{(1)}}&\equiv&Z^{(1)}=\det\aq\lap_1+\frac{R_0}{4}\cq\,
   \times\,\det\aq\beta\lap_1+\al\phantom{\frac{}{}}\cq^{1/2}
\nn\\ &&\hs\hs\hs\times\,
   \det\aq(\rho-3)\lap_0-R_0\phantom{\frac{}{}}\cq\,
   \times\,\det\aq \beta\at\lap_0+\frac{R_0}{4}\ct+\al\cq^{1/2}
\nn\\ &&\hs\hs\times\,
   \det\aq\at\lap_2-\frac{R_0}{6}\ct-\frac{X}{2f_0}\at\lap_2+\frac{R_0}3\ct\cq^{-1/2}\,
\nn\\ &&\hs\times\,
 \det\aq 2(\al+\beta\lap_1)\at\lap_1+\frac{R_0}{4}\ct+X\cq^{-1/2}
\nn\\ &&\times\,
  \det\ag\aq f''_0\at\lap_0+\frac{R_0}{3}\ct-\frac{2f_0}{3R_0}\cq
\,\aq \beta\at\lap_0+\frac{R_0}{4}\ct+\al\cq
\,\aq(\rho-3)\lap_0-R_0 \phantom{\frac{}{}}\cq^2
\cp\nn\\ &&\hs\hs\hs\hs\ap
+XC_1+X^2C_2
 \phantom{\frac{}{}}\cg^{-1/2}\,,
\label{PF}
\eeq
where $\Ga^{(1)}$ is the one-loop contribution
to the partition function and $C_1$ and $C_2$ are  operators, which read
\beq
C_1&=&{-\frac{2\,f\,R_0}{3}} + {\frac{2\,\al\,{R_0^2}}{3}}
+ {\frac{\beta\,{R_0^3}}{6}} +
  {\frac{f''_0\,{R_0^3}}{3}}
\nn\\&&\hs
+\at {-\frac{4\,f}{3}} + 3\,\al\,R_0 + {\frac{17\,\beta\,{R_0^2}}{12}} +
  {\frac{5\,f''_0\,{R_0^2}}{3}}
\cp
\nn\\ &&\hs\hs\hs\hs \ap
-   {\frac{2\,\al\,R_0\,\rho}{3}} -
  {\frac{\beta\,{R_0^2}\,\rho}{6}} -
  {\frac{\al\,R_0\,{{\rho}^2}}{3}} -
  {\frac{\beta\,{R_0^2}\,{{\rho}^2}}{12}}\ct\lap_0
\nn\\&&
+ \at 3\,\al + {\frac{15\,\beta\,R_0}{4}} + 2\,f''_0\,R_0 - 2\,\al\,\rho -
  {\frac{7\,\beta\,R_0\,\rho}{6}} + {\frac{\al\,{{\rho}^2}}{3}} -
  {\frac{\beta\,R_0\,{{\rho}^2}}{4}}\ct\lap_0^2
\nn \\ &&\hs\hs
+\frac{\beta}{3}(\rho-3)^2\lap_0^3\,,
\nn \\
C_2&=& \frac{2}{3}\,(\lap_0+R_0)\,. \label{AAA15} \eeq 
Equation
(\ref{PF}) reduces to the corresponding one in Ref.~\cite{frad},
when $f''=0$ (Einstein's gravity with a cosmological constant).
For another approach to the same problem see Ref~.\cite{gianni}.

In the derivation of (\ref{PF}), it is  understood that the
functional determinants has been  regularized by means of zeta
function regularization (see, for example \cite{eli94,byts96}).
However, we should remind that within the zeta function regularization, it is 
no longer true that 
\beq
\det AB=\det A \det B\,,
\eeq
where $A$ and $B$ are two (elliptic) operators. In fact, in general, one
has
\beq
\det AB=
e^{a(A,B)}\det A \det B\,,
\eeq
where $a(A,B)$ is a local functional called multiplicative anomaly
(see, for example, \cite{eli1,eli2}).
As a consequence, in the above manipulations, we have assumed the 
multiplicative anomaly to be trivial, namely $a(A,B)=0$. This is 
justified since
here we are limiting  ourselves to the one-loop approximation, and
in such a case, a non-trivial multiplicative anomaly, which is a local 
functional of the fields,  may be absorbed
into the renormalization ambiguity \cite{byts03}. 

Furthermore, another delicate point should be mentioned. The Euclidean 
gravitational action, due to the presence of $R$, is not bounded from below,
since  arbitrary negative contributions can be induced on $R$, by conformal 
rescaling of the metric. For this reason, 
we have also used the Hawking prescription of integrating over
imaginary scalar fields. Furthermore, the problem of the presence of
additional zero modes introduced by the decomposition (\ref{tt}) can
be treated  making use of the method presented in Ref.~\cite{frad}.

As one can easily verify, in the limit $X\to0$ (on-shell condition),
Eq.~(\ref{PF}) does not depend on the gauge parameters and reduces
to \beq \Ga_{on-shell}=I_E(g_0)+\Ga^{(1)}_{on-shell}&=&\frac{24\pi
f_0}{ G R_0^2}+ \frac12\,\ln\det\ \aq \ell^2
\at-\lap_2+\frac{R_0}6\ct\cq \nn\\&&
      -\frac12\,\ln\det \aq \ell^2 \at-\lap_1-\frac{R_0}4\ct \cq
\nn\\&&
    +\frac12\,
\ln\det \aq \ell^2 \at
-\lap_0-\frac{R_0}3+\frac{2f_0}{3R_0f''_0}\ct\cq\,. \label{AAA16}
\eeq As usual, an arbitrary renormalization parameter $\ell^2$ has
been introduced for dimensional reasons. When $f''_0=0$, namely in
the case of Einstein's gravity with a cosmological constant,
$f(R)=R-2\Lambda$, one obtains the well known result
\cite{perry,duff80,frad} \beq
\Ga_{on-shell}=I_E(g_0)+\Ga^{(1)}_{on-shell}&=&\frac{12\pi }{GR_0}
+\frac12\,\ln\det \aq \ell^2\at-\lap_2+\frac{R_0}6\ct\cq \nn \\&&
-\frac12\,\ln\det \aq \ell^2 \at-\lap_1-\frac{R_0}4\ct\cq\,.
\label{AAA17} \eeq

In order to simplify the off-shell computation, we choose the gauge
parameters $\rho=1,\beta=0$ and $\al=\ii$ ( Landau gauge). Thus, we
obtain
 \beq
\Ga &=&\frac{24\pi}{GR_0^2}\,f_0
+\frac12\ln\det\at-\lap_2-\frac{R_0}{6}\,\,\frac{X+2f_0}{X-2f_0}\ct
\nn \\&&
       -\frac12\ln\det\at-\lap_1-\frac{R_0}{4}\ct
        -\frac12\ln\det\at-\lap_0-\frac{R_0}{2}\ct
\nn\\ &&
   +\frac12\ln\det\ag\at-\lap_0
    -\frac{5R_0}{12}-\frac{X-2f_0}{6R_0f''_0}\ct^2\,
\cp\nn\\ &&\ap\hs\hs\hs
   -\aq\at\frac{5R_0}{12}+\frac{X-2f_0}{6R_0f''_0}\ct^2
-\frac{R^2_0}{6}-\frac{X-f_0}{3f''_0} \,\cq\cg\,. \label{EA-G}\eeq
Recall now that the functional determinant of a differential
operator $A$ can be defined in terms of its zeta function by means
of (see for example \cite{eli94,byts96}) \beq
\zeta(s|A)=\sum\la_n^{-s}\,,\hs\hs \Re s>\frac{D}2\,,
\label{ZFdef}\eeq \beq
\ln\det(\ell^2A)=-\zeta'(0|\ell^2A)=-\zeta'(0|A)+\ln\ell^2\zeta(0|A)\,,
\label{logDET}\eeq 
where the prime indicates derivation with respect
to $s$. 
Looking at Eq.~(\ref{EA-G}), we see that the one-loop
effective action can be written in terms of the derivative of zeta
functions corresponding to Laplace-like operators acting on scalar,
vector and tensor fields on a 4-dimensional de Sitter space. In all
such cases, the eigenvalues of the Laplace operator are explicitly
known and the zeta-functions can be computed directly using
Eq.~(\ref{ZFdef}). For the reader's convenience, we have reported in
the Appendix \ref{AppX} all the details of the method used in the
explicit computation for the example that will follow.

Finally, equations (\ref{EA-G}), (\ref{logDET}) and (\ref{Q0}),
(\ref{Q1}), and (\ref{Q2}) in Appendix \ref{AppX} lead to the
off-shell one-loop effective action \beq
\label{331}
\Ga=&\frac{24\pi}{GR_0^2}\,f_0
-\frac12\,Q_2(\al_2)+\frac12\,Q_1(\al_1)
+\frac12\,Q_0(\al_0)-\frac12\,Q_0(\al_+) -\frac12\,Q_0(\al_-)\,,
\label{Gamma}\eeq where (see App.~\ref{AppX}) \beq
\al_2&=&\frac{17}4+q_2\,,\hs\hs\hs q_2=2\,\,\frac{X+2f_0}{X-2f_0}\,,
\\
\al_1&=&\frac{13}{4}+q_1=\frac{25}{4}\,,\hs\hs q_1=3\,,
\\
\al_0&=&\frac{9}{4}+q_0=\frac{33}{4}\,,\hs\hs q_0=6\,,
\\
\al_\pm&=&\frac{9}{4}+q_\pm\,,\hs\hs
q_\pm=5+2\,\,\frac{X-2f_0}{R_0^2f''_0}
\nn\\ &&\hs\hs\hs\hs\hs
    \pm\sqrt{\at 5+2\,\frac{X-2f_0}{R_0^2f''_0}\ct^2
         -24\,\at1+2\,\frac{X-f_0}{R^2_0f''_0}\ct}\,.
\label{AAA18}
\eeq

Now, we would like to present the explicit example for the model described by
Eq.~(\ref{invR12}). First of all we consider the simplest case in
the class of models defined by Eq.~(\ref{invR12}),  thus \beq
f(R)=R-\frac{\mu_1}{R}-\mu_2\,, \hs X=R_0-3\frac{\mu_1}{R_0}-2
\mu_2\, \label{AAA19} \eeq We may eliminate $X$ and get \beq
\al_2&=&\frac{57\mu_1+R_0(32\mu_2-7R_0)}{4(\mu_1+R_0^2)}\,,
\\
\al_\pm&=&\frac{33}{4}+\frac{R_0^2}{\mu_1}\pm\frac{1}{\mu_1}
\sqrt{R_0^4-36\mu_1^2+12\mu_1R_0(R_0-2\mu_2)}\,.
\label{AAA20}
\eeq
Hence, the one-loop effective action in $f(R)$ gravity in de Sitter space 
is found.

In the next section the above effective action will be applied 
to study the back-reaction of $f(R)$ gravity to background geometry.
However, several important remarks are in order. As usually, any perturbative 
calculation of the effective action in quantum gravity is gauge dependent.
The way to resolve such a problem is well-known: to use 
the gauge-fixing independent effective action (for a review, see \cite{buch}).
 More serious problem is related with the fact that quantum gravity
 under investigation is not renormalizable. Then, generally speaking,
 higher order corrections are of the same order as one-loop ones (the same 
is applied to all previous quantum considerations of Einstein gravity).
As a result all one-loop conclusions are highly questionable as they 
maybe spoiled by higher loops effects.
In this respect, the results of our work are definitely useful in
 the following sense.
One can expect that perturbatively renormalizable gravity maybe constructed 
for some version of $f(R)$ gravity. (So far only higher derivative gravity 
is known to be renormalizable). In this case, our work gives necessary 
background for one-loop quantization of such theory. From another side, 
to get the meaningful results with non-renormalizable quantum gravity 
one may apply the exact renormalization group scheme. In such a case, 
higher loop effects are not important. Our work maybe considered also as 
necessary and important step in this direction. Indeed, it is  
technically clear how   to construct the exact RG equations for $f(R)$ 
gravity using results of this section in the analogy with Einstein gravity
\cite{falkenberg}.

\section{Quantum-corrected de Sitter cosmology}

Let us consider the role of quantum effects to the background
cosmology. So far, such study has been done for Einstein or higher
derivatives gravity only. In order to see the difference with such
models, we take the example of modified gravity with the action \beq
\label{I} f(R)=R - \frac{\mu_1}{R}\,. \eeq It is interesting to
investigate the region where curvature is not very big, as otherwise
the classical theory is effectively reduced to Einstein's gravity.
Moreover, if curvature is small one can neglect the powers of
curvature in the one-loop effective action supposing that
logarithmic terms give the dominant contribution. The parameter
$\mu_1$ is chosen to be very small in order to avoid conflicts with
Newton's law. As a result, one obtains \be \label{II} \Gamma (R_0) =
\frac{24\pi}{GR_0^2}\left( R_0 - \frac{\mu_1}{R_0}\right) +
\left(\alpha + \frac{\beta}{\mu_1 + R_0^2}\right) \ln \left(
\frac{l^2 R_0}{12} \right)\ . \ee Here $\alpha$ and $\beta$ are
constants. It is assumed that the curvature is constant, $R=R_0$.
Let us find the minimum of $\Gamma$ with respect to $R_0$. One can
write $\Gamma'\left(R_0\right)$ as \beq \label{III}
\Gamma'\left(R_0\right) &=& F\left(R_0\right) - G\left(R_0\right)\ , \nn\\
F\left(R_0\right) &\equiv & \frac{24\pi}{G R_0^2} \left( -1 + \frac{3\mu_1}{R_0^2}\right) \ ,\nn\\
G\left(R_0\right) &\equiv & \frac{2\beta R_0}{\left(\mu_1 + R_0^2\right)^2}
\ln \left( \frac{l^2 R_0}{12} \right)
 - \frac{1}{R_0}\left(\alpha + \frac{\beta}{\mu + R_0^2}\right)\ .
\eeq When \be \label{IV} R_0=R_c\equiv \sqrt{3\mu_1}\ , \ee
$F(R_0)=0$, which corresponds to the classical solution. When
$R_0\sim 0$, $F\left(R_0\right)$ behaves as $F\left(R_0\right)\sim
\frac{72\pi\mu_1}{G R_0^4}$, and when $R_0\to +\infty$,
$F\left(R_0\right)\to -\frac{24\pi}{G R_0^2}$. Since \be \label{V}
F' \left( R_0 \right)  =  \frac{48\pi}{G R_0^3} \left( 1 -
\frac{6\mu_1}{R_0^2}\right)\ , \ee there is a minimum for $F \left(
R_0 \right)$ when $R_0=\sqrt{6\mu_1}> R_c$.
On the other hand, if $\alpha\neq 0$, $G(R_0)$ behaves as
\be
\label{VI}
G(R_0)\to - \frac{1}{R_0}\left(\alpha + \frac{\beta}{\mu_1}\right)\ ,
\ee
when $R_0\to 0$ and
\be
\label{VII}
G(R_0)\to - \frac{\alpha}{R_0}\ ,
\ee
when $R_0\to + \infty$.
Hence for $R_0>0$, if $\alpha<0$, $\Gamma'(R_0)>0$
when $R_0\to +0$ and $\Gamma'(R_0)<0$ when
$R_0\to +\infty$.
Therefore, there is a solution which satisfies $\Gamma'(R_0)=0$ if $\alpha<0$.
When $\alpha>0$, the existence of the solution depends on the details of the
parameters.

In case $\alpha=0$, when $R_0\to 0$, $G(R_0)$ behaves as \be
\label{VIII} G(R_0)\to - \frac{\beta}{\mu_1 R_0}\ , \ee and when
$R_0\to + \infty$, we find \be \label{IX} G(R_0)\to
\frac{2\beta}{R_0^3}\ln \left( \frac{l^2 R_0}{12} \right) \ . \ee
When $R_0>0$, if $\beta>0$, $\Gamma'(R_0)>0$ when $R_0\to +0$ and
$\Gamma'(R_0)<0$ when $R_0\to +\infty$. Hence even if $\alpha=0$,
when $\beta>0$, there is a solution for equation $\Gamma'(R_0)=0$.
When $\beta<0$, the existence of the solution depends on the details
of the parameters again. Thus, there could be a positive non-trivial
solution for $R_0$, which describes the quantum-corrected de Sitter
space. One may play with the parameters of the theory under
consideration in such a way that the quantum-corrected de Sitter
space can provide a solution to the cosmological constant problem.
The above results indicate that the classical de Sitter solution
(\ref{IV}) can survive when one takes into account the quantum
corrections.  A similar consideration can be done for any specific
$f(R)$ gravity.

Let us demonstrate that indeed with some fine-tuning the obtained 
effective action maybe used to resolve the cosmological constant problem. 
One can present (\ref{331}) corresponding to (\ref{I}) as 
\be 
\label{C1} 
\Gamma = \frac{24\pi}{GR_0^2}\left(R_0 - \frac{\mu_1}{R_0}\right) 
+ Q\left(l^2; R_0, \mu_1\right)\ . 
\ee 
In general $Q\left(l^2; R_0, \mu_1\right)$ has a structure as 
\be 
\label{C1b} 
Q\left(l^2; R_0, \mu_1\right) 
= Q_0\left(\frac{R_0^2}{\mu_1}\right) \ln \frac{l^2R_0}{12} 
+Q_1\left(\frac{R_0^2}{\mu_1}\right)\ . 
\ee 
By the condition that $\Gamma$ takes a minimum value with the variation 
over $R_0$, 
we obtain 
\be 
\label{C2} 
0=\frac{\partial \Gamma}{\partial R_0} = \frac{24\pi}{G}\left(- \frac{1}{R_0^2} 
+ \frac{3\mu_1}{R_0^4}\right) 
+ \frac{\partial Q\left(l^2; R_0, \mu_1\right)}{\partial R_0}\ . 
\ee 
The convenient choice between the parameters is 
\be 
\label{C3} 
\left(\frac{12}{l^2}\right)^2=c_0^2 \mu_1\ . 
\ee 
Here $c_0$ is a constant which could be determined later. 
Then $Q$ has the following form: 
\be 
\label{C4} 
Q=Q\left(\frac{R_0^2}{\mu_1}\right) 
= Q_0\left(\frac{R_0^2}{\mu_1}\right) \ln \left(\frac{1}{c_0}\sqrt{\frac{R_0^2}{\mu_1}}\right) 
+Q_1\left(\frac{R_0^2}{\mu_1}\right) 
\ . 
\ee 
We now consider the possibility that  the vanishing cosmological constant 
could be obtained 
by (fine-) tuning the parameters. 
The corresponding condition that the vacuum energy, or cosmological 
constant, vanishes 
requires 
\be 
\label{C5} 
\Gamma=0\ , 
\ee 
which may be solved with respect to $\mu_1$ as $\mu_1=\mu_1\left(R_0\right)$, 
  what gives 
\be 
\label{C6} 
0=\frac{\partial \Gamma}{\partial R_0} + \frac{d\mu_1}{dR_0}\frac{\partial \Gamma}{\partial \mu_0}\ . 
\ee 
By combining (\ref{C2}) and (\ref{C6}) with (\ref{C4}), one gets 
\be 
\label{C7} 
0= \frac{\partial \Gamma}{\partial \mu_0} 
= \frac{24\pi}{GR_0^3} - \frac{R_0^2}{\mu_1^2}Q'\left(\frac{R_0^2}{\mu_1}\right)\ , 
\ee 
which gives 
\be 
\label{C8} 
Q'\left(\frac{R_0^2}{\mu_1}\right)= - \frac{24\pi\mu_1^2}{GR_0^5}\ . 
\ee 
Then by using (\ref{C2}), (\ref{C4}), and (\ref{C8}), it follows 
\be 
\label{C9} 
0 = \frac{24\pi}{G}\left(- \frac{1}{R_0^2} 
+ \frac{3\mu_1}{R_0^4}\right) + \frac{2R_0}{\mu_1}Q'\left(\frac{R_0^2}{\mu_1}\right) 
= \frac{24\pi}{G}\left(- \frac{1}{R_0^2} + \frac{\mu_1}{R_0^4}\right) \ . 
\ee 
Hence, 
\be 
\label{C10} 
R_0^2 = \mu_1\ . 
\ee 
By using (\ref{C1}), (\ref{C4}), and (\ref{C10}), we find 
\be 
\label{C11} 
Q(1)=0\ . 
\ee 
Then Eq.(\ref{C4}) shows that 
\be 
\label{C12} 
c_0=\e^{-\frac{Q_1(1)}{Q_0(1)}}\ . 
\ee 
Therefore, including the quantum corrections and (fine-)tuning the 
theory parameters, we may obtain 
the solution expressing the vanishing (effective) cosmological constant. 
Of course, such solution of cosmological constant problem is one-loop, 
and in higher orders better fine-tuning maybe required.

The effective action (\ref{II}) has been evaluated in the Euclidean
signature, in which case we should  recall that the 4d de Sitter
space with positive constant curvature $R_0$ becomes a sphere of
radius \be \label{X} a=\sqrt{\frac{12}{R_0}}\ . \ee The volume
(area) of the sphere $V$ is \be \label{XI} V=\frac{8\pi^2
a^4}{3}=\frac{384\pi^2}{R_0^2}\ . \ee Identifying $\int d^4x
\sqrt{g} \sim V = {384\pi^2}{R_0^2}$, one may reasonably assume the
local effective Lagrangian corresponding to (\ref{II}) to be \bea
\label{XII}
\Gamma &=& \frac{1}{384\pi^2}\int \sqrt{g}L_{\rm eff}(R)\ ,\\
L_{\rm eff}(R)&=& \frac{24\pi}{G} \left( R - \frac{\mu_1}{R}\right)
+ R^2 \left(\alpha + \frac{\beta}{\mu_1 + R^2}\right) \ln \left(
\frac{l^2 R}{12} \right)\, . \nonumber \eea The effective equation
of motion is  \be \label{XIII} 0=\frac{1}{2}g_{\mu\nu} L_{\rm
eff}(R) - R_{\mu\nu}L_{\rm eff}'(R) + \nabla_\mu \nabla_\nu L_{\rm
eff}'(R) - g_{\mu\nu}\nabla^2 L_{\rm eff}'(R) \, , \ee with the
curvature being covariantly constant, $\nabla_\rho R_{\mu\nu}=0$,
Eq.~(\ref{XIII}) reduces to $\Gamma'(R_0)=0$ in (\ref{III}).
Supposing the FRW metric with flat 3-dimensional part, \be
\label{XIV} ds^2 = - dt^2 +
\e^{2a(t)}\sum_{i=1,2,3}\left(dx^i\right)^2\ , \ee the
$(t,t)$-component of (\ref{XIII}) has the following form \beq
\label{XV} 0&=&-\frac{1}{2}L_{\rm eff}\left(6\dot H + 12 H^2\right)
+ 3\left(\dot H + H^2\right) L_{\rm eff}'\left(6\dot H + 12
H^2\right)
\nn\\
&& - 3H\frac{d}{dt}\left( L_{\rm eff}'\left(6\dot H + 12
H^2\right)\right)\ . \eeq The Hubble parameter $H$ is defined by
$H\equiv \frac{\dot a}{a}$, as usual. We now split $L_{\rm
eff}(R)=L_c(R) + L_q(R)$, with \beq \label{XVb} L_c(R) &\equiv&
\frac{24\pi}{G} \left( \frac{R^3}{4\mu_1} -
\frac{R}{2} + \frac{5\mu_1}{4R}\right) \ ,\nn \\
L_q(R) &\equiv& R^2 \left(\alpha + \frac{\beta}{\mu_1 + R^2}\right)
\ln \left( \frac{l^2 R}{12} \right) \ . \eeq Let us assume that
$L_q(R)$ are much smaller than $L_c(R)$ and consider the
perturbation from the classical solution in (\ref{IV}), by putting
\be \label{XVI} H=h_c + \delta h\ , \quad h_c\equiv
\sqrt{\frac{R_c}{12}}=\sqrt{\frac{\sqrt{3\mu_1}}{12}}\ , \ee or \be
\label{XVII} R=R_c + \delta R\ , \quad \delta R\equiv 6\delta \dot h
+ 24 h_c \delta h\ . \ee Note that  $L_c(R)$ contains the quantum
correction. From (\ref{XV}) it follows that \beq \label{XVIII} 0&=&
- 18h_c L''_{c0} \delta \ddot h - 54 h_c^2 L''_{c0} \delta \dot h +
\left( - 6 h_c L'_{c0} + 72 h_c^3 L''_{c0}\right) \delta h
 - \frac{1}{2}L_{q0} + 3h_0^2 L'_{q0}
\nn\\
&=& \frac{24\pi}{G h_c} \delta \ddot h + \frac{72\pi}{G} \delta \dot h
 - \frac{288 \pi h_c}{G}\delta h
\nn\\
&& + 24h_c^2 \left(\alpha - \frac{\beta}{768 \mu_1}\right)\ln \left(l^3 h_c^2\right)
+ 12h_c^2 \left(\alpha + \frac{\beta}{192\mu_1}\right)
\nn\\
&=& \frac{24\pi}{G h_c} \left[ \delta \ddot h + 3h_c  \delta \dot h - 12 h_c^2 \delta h \right. \nn\\
&& \left. - \frac{G h_c} {2\pi}\left\{ 2\left(\alpha -
\frac{\beta}{768 \mu_1}\right)\ln \left(l^3 h_c^2\right) +
\left(\alpha + \frac{\beta}{192\mu_1}\right)\right\} \right] \ .
\eeq Here $L'_{c0}=L'_c\left(R_0\right)$ and
$L''_{c0}=L''_c\left(R_0\right)$. Then the solution is \be
\label{XIX} \delta h = h_1 + A_+ \e^{\alpha_+ t} + A_- \e^{\beta_-
t}\ , \ee with \beq \label{XX} h_1&=& \frac{G h_c} {24\pi}\left\{
2\left(\alpha - \frac{\beta}{768 \mu_1}\right)\ln \left(l^3
h_c^2\right)
+ \left(\alpha + \frac{\beta}{192\mu_1}\right)\right\}\ ,\nn\\
A_\pm &\equiv& \frac{1}{2}\left( -3\pm \sqrt{57} \right) h_c\ . \eeq
Since $A_+>0$, the de Sitter solution becomes unstable under the
perturbation.

Thus, for a specific modified gravity model we have demonstrated
that the quantum gravity correction shifts the radius of the de
Sitter space trying to destabilize the de Sitter phase. This may
find interesting applications in the study of the issue of the exit
from inflation or in the study of the decay of the dark energy
phase.

\section{Black hole nucleation rate}

We have remarked (see Appendix \ref{solution}) that within the
modified gravitational models we are dealing with, there is room for
black hole solutions, formally equivalent to black hole solutions of
the  Einstein theory with a non vanishing cosmological constant. As
in the Einstein case, one is confronted with  the black hole
nucleation problem \cite{ginspar}. We review here the  discussion
reported in refs. \cite{ginspar,wipf}.

To begin with, we recall that we shall deal with a tunneling process in 
quantum gravity. On general backgrounds, this process is mediated by the
associated gravitational instantons, namely stationary solutions of Euclidean
gravitational action, which dominate the path integral of Euclidean quantum
gravity. It is a well known fact that as soon as an imaginary part appears in 
the one-loop partition 
function, one has a metastable thermal state and thus 
a non vanishing decay rate.
Tipically, this imaginary part comes from the existence of a negative mode
in the one-loop functional determinant.
Here, the semiclassical and one-loop approximations are
the only techniques at disposal, even though one should bear in mind their
limitations as well as their merits.

Let us consider  a general model described 
by $f(R)$ with $\Lambda_{eff} >0$. Thus, we may have de Sitter and Nariai Euclidean instantons. 
Making use of the  instanton approach,
 we have for the Euclidean partition function \be
Z\simeq Z(S_4)+Z(S_2 \times S_2)= Z^{(1)}(S_4)e^{-I(S_4)}+Z^{(1)}
(S_2 \times S_2)e^{-I(S_2 \times S_2)}\,, \ee where $I$ is the
classical action and $Z^{(1)}$  the quantum correction, typically a
ratio of functional determinants. The classical action can be easily
evaluated and reads \be I(S_4)=-\frac{24f_0}{GR^2_0}\,,\hs I(S_2
\times S_2)=-\frac{16f_0}{GR^2_0}\,. \ee

At this point, we make a brief  disgression regarding the entropy of
the above black hole solutions. To this aim, we follow the arguments
reported in Ref.~\cite{brevik}. If one make use of the Noether
charge method for evaluating the entropy associated with black hole
solutions with constant curvature in modified gravity models, one
has \be S=\frac{A_H}{4G} f'(R_0)\,. \label{ve} \ee As a consequence,
in general, one obtains a modification of the ``Area Law''. 
For stable models like
 (\ref{q}) with $p>0$ (see below), one has $f'_0=1+8p\Lambda$. 
For the model (\ref{invR}),  
$f'_0=\frac{4}{3}$, and
thus \cite{brevik}, \be S=\frac{A_H}{3G} \,. \label{ve1} \ee 
In the
above equations, $A_H=4\pi r_H^2$,  $r_H$ being the radius of the
event horizon or cosmological horizon related to a black hole
solution. This turns out to be model dependent.  
It is interesting to note that for unstable modified gravity 
(with negative first derivative of $f$) the entropy may be negative!

We are interested
in the case of de Sitter space, and we have \be A_H=\frac{12
\pi}{\Lambda_{eff}}=\frac{48 \pi}{R_0}\,. \ee Thus, for the de
Sitter solution,  \be S(S_4)=\frac{12 \pi}{G R_0} f'(R_0)\,
\label{ve2} \ee and if we take into account the Eq.~(\ref{B}), one
gets \be S(S_4)=-I(S_4)\,. \ee 
With the help of the one-loop effective action one can calculate
quantum correction for classical entropy.

We may introduce the free energy
$F=-\frac{S}{\beta}$, where $\beta=2\pi \sqrt{\frac{12}{R_0}}$ is
the inverse of the Hawking temperature for the de Sitter space. As a
consequence, we have \cite{wipf} \be Z\simeq  Z^{(1)}(S_4)e^{-\beta
\hat F}, \ee where \be \hat F=F(S_4)-\frac{Z((S_2 \times S_2)}{\beta
Z(S_4)}\,. \ee The rate of quasiclassical decay in the de Sitter
space is present as soon as $\hat F$ has a non vanishing imaginary
part and it is given by $N= 2 \Im \hat F$. When $f(R)=R-2\Lambda$,
the Einstein case, it turns out that $Z((S_2 \times S_2)$ has an
imaginary part but $Z(S_4)$ is real. As a result, in the Einstein
case, the nucleation rate is \cite{ginspar,wipf} \be N=-2 \frac{\Im
Z((S_2 \times S_2))}{\beta Z(S_4)}\,. \ee Within our generalized
models, the dynamics of the gravitational field is different. In
fact, also in the de Sitter case, due to the presence of an
additional term in the on-shell one-loop effective action related to
the operator $L_0$, see  Eq. (\ref{AAA16}), there exists the
possibility of  negative modes. In fact, from Appendix B, one has
\be \lambda_n(L_0)=\frac{R_0}{12}\at n^2+3n-4+\frac{4f_0'}{R_0f_0''}
\ct\,, \ee where $n=0,1,2,3,..$. It is clear that we have negative
modes as soon as \be \frac{4f_0'}{R_0f_0''}<0 \,. \label{q1} \ee For
example, for the model \be f(R)=R-\frac{\mu}{R^n}\,, \ee the
quantity (\ref{q1}) is always negative and one obtains, at least, two
negatives modes.

For the model \be f(R)=R+pR^2-2\Lambda\,, \ee 
there are no negative
modes as soon as $p>0$, in agreement with the classical stability
observed in ref.~\cite{barrow83}. 
For $p<0$, one has only a negative mode
when 
\be p<-\frac{1}{8 \Lambda}\,. \ee 
Thus, in general, within
these specific modified gravitational models, de Sitter space is  unstable
due to quantum corrections.

\section{Discussion}

In summary, we have here calculated the one-loop effective action for 
general $f(R)$ gravity in the de Sitter space. Generalized zeta 
regularization has been used to obtain a finite answer for the 
functional determinants in the effective action, what has proven to 
be a very convenient procedure. The important lesson to be drawn 
from this calculation, generalizing the previous 
program for one-loop Einstein gravity in the de Sitter background, 
is that quantum corrections tend to destabilize the classical de 
Sitter universe.
The constant curvature black hole nucleation rate can be discussed
 within our scheme. Note that, typically, $f(R)$ models  may contain 
(depending of course on the explicit form of $f$)
more negative modes, aside from the Einstein one.

Another lesson we have learned is that the inverse powers of 
the curvature (which are important at current epoch) 
do not arise from the perturbative quantum corrections,
as has been explicitly demonstrated here. Some remarks about black 
holes and associated entropy in modified gravity have been also made. 

There are several directions in which our results can be extended
and applied. First of all, one can repeat the whole calculation 
of the functional determinants for the case of Anti-de Sitter space. 
Then, the one-loop effective action  found in the third section can be
given also for the Anti-de Sitter universe.
In relation with the AdS/CFT correspondence, our approach
can  be very interesting for the study of the (de)stabilization of 
such a universe against quantum corrections. On the other hand, 
it can be also important in the search for the solution of the 
cosmological constant problem within hyperbolic backgrounds 
in the large distance limit \cite{BOZ}, what will be discussed 
elsewhere. Second, having in mind the current interest on
gauged supergravities in  de Sitter space, one can try to 
formulate the theory of $f(R)$ gauged supergravity and to 
study the relevance of quantum effects for such theories.
\medskip

\section*{Acknowledgments}
We would like to thank L. Vanzo for useful discussions. Support from
the program INFN(Italy)-CICYT(Spain), from DGICYT (Spain), project
BFM2003-00620 and SEEU grant PR2004-0126 (EE), is gratefully
acknowledged.
\medskip

\appendix

\section{Black hole solutions in modified gravity}
\label{solution}

Here we revisit the class of general static neutral black hole
solutions in four dimensions and non vanishing cosmological
constant. Let us start with the usual Ansatz for the metric \be
ds^2=A(r)dt^2-\frac{dr^2}{A(r)}-r^2d\Sigma_k^2\, , \label{ds} \ee
where $k=0,\pm 1$ and the horizon manifolds are $\Sigma_1=S^2$, the
two dimensional sphere, $\Sigma_0=T^2$, the two dimensional torus,
and $\Sigma_{-1}=H^2/\Gamma$, the two dimensional compact Riemann
surface. The scalar curvature and the non vanishing components of
the Ricci tensor read \be R=-\frac{1}{r^2}\aq r^2A''+4rA'+2A-2k
\cq\,, \ee \be R_{tt}=-\frac{1}{2r}\aq rA''+2A'\,, \cq \ee \be
R_{rr}=\frac{1}{2r A}\aq rA''+2A'\,, \cq \ee \be R_{ab}= g_{ab}\aq
k- rA'-A \,. \cq \ee If we look for a constant curvature solution,
we should have \be r^2A''+4rA'+2A-2k=-r^2 R_0 \,. \ee The general
exterior solution depends on two integration constants, $c$ and $b$,
and reads \be A(r)=\frac{b}{r^2}+k-\frac{c}{r}-\frac{R_0}{12}r^2\,.
\ee However, this solution has to satisfy the equation of motion \be
R_{\mu\nu}=\frac{R_0}{4}g_{\mu\nu}\,. \label{E1} \ee As a result, it
is easy to show that these equations are  satisfied with $b=0$ and
$c$ arbitrary, and we have (see for instance \cite{vanzo,brevik})
\be A(r)=k-\frac{c}{r}-\frac{R_0}{12}r^2\,. \ee Since $A(r)>0$, we
can have $k=1$ only for positive $R_0$, and this is the
Schwarzschild-de Sitter solution. For $R_0 <0$, we may have $k=1$,
namely the Schwarzschild-AdS  solution. We may also have  $k=0$,
with a torus topology for the horizon manifold, and $k=-1$, with an
hyperbolic topology for the horizon topology, the so called
topological black holes \cite{vanzo}. The constant $c$ is related to
the mass of the black hole. The de Sitter solution is obtained when
$R_0 >0$ and with $c=0$ and $k=1$, the AdS solution is obtained when
$R_0 <0$ and with $c=0$ and $k=1$. For $c$ non vanishing, one has
black hole solutions. These black hole solutions may have extremal
cases and extremal limits. The extremal case exists for $k=-1$. For
$k=1$, $R_0 >0$, one has only the extremal limit of the
Schwarzschild-de Sitter solution (see, for example  \cite{calda} and
references therein), and the metric reduces to \be
ds^2=\frac{4}{R_0}\at dS_2^2+ dS_2^2 \ct \,. \ee This is a  space
with constant curvature $R_0$ solution of Eqs.~(\ref{E1}).

\section{Evaluation of functional determinants}
\label{AppX} Here we shall make use of zeta function regularization
for the evaluation of the functional determinants appearing in the
one-loop effective action, Eq.~(\ref{EA-G}) computed in the previous
section.

First,  we  outline the standard technique, based on binomial
expansion, which relates the zeta-functions corresponding to the
operators $\hat A$, with eigenvalues $\hat\la_n>0$ and
$A=\frac{R_0}{12}(\hat A-\al)$, with eigenvalues
$\la_n=\frac{R_0}{12}(\hat\la_n-\al)$, $\al$ being a real constant.
With this choice, $\hat\la_n$ and $\al$ are dimensionless. We assume
to be dealing with a second-order differential operator on a $D$
dimensional compact manifold. Then, by definition, for $\Re s>D/2$
one has \beq \hat\ze(s)&\equiv&\ze(s|\hat A)=\sum_n\hat\la_n^{-s}\,,
\label{sum1}
\\
\ze_\al(s)&\equiv&\ze(s|A)=\sum_n\la_n^{-s}=\at
\frac{R_0}{12}\ct^{-s} \sum_n(\hat\la_n-\al)^{-s}\,,
\label{sum2}\eeq where, as usual, zero eigenvalues have to be
excluded in the sum. In order to use the binomial expansion in
(\ref{sum2}), we have to treat separately the several terms
satisfying the condition $|\hat\la_n|\geq|\al|$. So, we write \beq
\ze_\al(s)=\at \frac{R_0}{12}\ct^{-s}\aq
F_\al(s)+\sum_{k=0}^\ii\frac{\al^k\Ga(s+k)\hat G(s+k)}{k!\Ga(s)}\cq
\,, \label{binE}\eeq where we have set \beq
F_\al(s)&=&\sum_{\hat\la_n\leq|\al|;\,\,\hat\la_n\neq\al}\,(\hat\la_n-\al)^{-s}\,,
\hs\hs \hat F(s)=\sum_{\hat\la_n\leq|\al|}\,\hat\la_n^{-s}\,,
\\
\hat G(s)&=&\sum_{\hat\la_n>|\al|}\,\hat\la_n^{-s} =\hat\ze(s)-\hat
F(s)\,,\hs\hs F_\al(0)-\hat F(0)=N_0\,, \label{BBB1} \eeq $N_0$
being the number of zero-modes. It has to be noted that (\ref{binE})
is valid also in the presence of zero-modes or negative eigenvalues
for the operator $A$. In many interesting cases, $F_\al(s)$ and
$\hat F(s)$ are vanishing and thus $\hat G(s)=\hat\ze(s)$.

As is well known, the zeta function has simple poles on the real
axis for $s\leq D/2$ but it is regular at the origin. Of course, the
same analytic structure is also valid for the function $\hat G(s)$.
One has \beq \Ga(s)\hat\ze(s)=\sum_{n=0}^\ii\,\frac{\hat
K_n}{s+(n-D)/2}+\hat J(s)\,, \label{BBB2} \eeq $\hat J(s)$ being an
analytic function and $\hat K_n$ the heat-kernel coefficients
depending on geometrical invariants. In the physical applications we
have to consider, we have to deal with the zeta function and its
derivative at zero, thus it is convenient to consider the Laurent
expansion around $s=0$ of the functions \beq
\Ga(s+k)\hat\ze(s+k)&=&\frac{\hat b_k}{s}+\hat a_k+O(s)\,,
\\
\Ga(s+k)\hat G(s+k)&=&\frac{b_k}{s}+a_k+O(s)\,,
\label{BBB3}
\eeq
\beq
b_0&=&\hat b_k-\hat F(0)\,,
\hs\hs
a_0=\hat a_0+\ga\hat F(0)\,,\\
b_k&=&\hat b_k=\hat K_{D-2k}\,,
\hs\hs a_k=\hat a_k-\Ga(k)\hat F(k)\,,
\hs\hs 1\leq k\leq\frac{D}2\,,\\
b_k&=&\hat b_k=0\,,
\hs\hs
\hat G(k)=\hat\ze(k)-\hat F(k)\,,
\hs\hs k>\frac{D}2\,.
\label{BBB4}
\eeq

Now, from previous considerations, one  obtains \beq
\ze_\al(s)&=&\at \frac{R_0}{12}\ct^{-s} \aq\sum_{0\leq k\leq D/2}\,
\at\frac{b_k\al^k}{k!} +s\,\frac{(a_k+\ga b_k)\al^k}{k!}\ct \cp\nn\\
&& \ap\hs\hs\hs +F_\al(s)+s\sum_{k>D/2}\,\frac{\al^k\hat
G(k)}{k}+O(s^2)\cq \,,\label{BBB5} \eeq and finally \beq
\ze_\al(0)&=&F_\al(0)+\sum_{0\leq k\leq D/2}\,\frac{
b_k\al^k}{k!}\,,
\\
\ze'_\al(0)&=&-\ze_\al(0)\,\ln\frac{R_0}{12}+\sum_{0\leq k\leq D/2}\,
\frac{(a_k+\ga b_k)\al^k}{k!}
+F_\al'(0)+\sum_{k>D/2}\,\frac{\al^k\hat G(k)}{k}\,,
\label{ze1}\eeq
$\ga$ being the Euler-Mascheroni constant.
If there are negative eigenvalues then
$F'_\al(0)$ has an imaginary part, which reflects instability of the model.

In the paper we have to deal with Laplace-like operators acting on scalar
and constrained vector and tensor fields in a 4-dimensional de Sitter
space $SO(4)$.
In all such cases, the eigenvalues $\la_n$ and relative degeneracies $g_n$
can be written in the form
\beq
\la_n=\frac{R_0}{12}\at\hat\la_n-\al\ct\,,
\hs g_n=c_1\,\at n+\nu\ct+c_3\,\at n+\nu\ct^3\,,
\hs\hat\la_n=\at n+\nu\ct^2\,,
\label{hlan}\eeq
where $n=0,1,2...$ and
$c_1,c_2,\nu,\al$ depend on the operator one is dealing with. In our cases
 we have
\beq
L_0&=&-\lap_0-\frac{R_0}{12}\,q\segue\ag
\begin{array}{ll}
  \nu=\frac32\,,&
   \al=\frac{9}{4}+q\,,\\ \\
   c_1=-\frac{1}{12}\,,& c_3=\frac13\,.
\end{array}\cp
\label{L0PP}
\\
L_1&=&-\lap_1-\frac{R_0}{12}\,q\segue\ag
\begin{array}{ll}
  \nu=\frac52\,,&
   \al=\frac{13}{4}+q\,,\\ \\
   c_1=-\frac{9}{4}\,,&  c_3=1\,.
\end{array}\cp
\label{L1PP}
\\
L_2&=&-\lap_2-\frac{R_0}{12}\,q\segue\ag
\begin{array}{ll}
   \nu=\frac72\,,&
   \al=\frac{17}{4}+q\,,\\ \\
   c_1=-\frac{125}{12}\,,& c_3=\frac53\,,
\end{array}\cp
\label{L2PP}\eeq
where $q$ are dimensionless parameters depending on the
specific choice of $f(R)$.

We note that $\hat\ze(s)$ is related to well known
Hurwitz functions $\ze_H(s,\nu)$ by
\beq
\hat\ze(s)&=&\sum_{n=0}^\ii\,g_n\hat\la_n^{-s}=
\sum_{n=0}^\ii\,\aq c_1\at n+\nu\ct^{2s-1}
+c_3\at n+\nu\ct^{2s-3}\cq
\nn\\
&=&c_1\ze_H\at 2s-1,\nu\ct+
c_3\ze_H\at 2s-3,\nu\ct
\label{BBB6}
\eeq
and
\beq
\hat G(s)&=&c_1\ze_H\at 2s-1,\nu\ct+c_3\ze_H\at 2s-3,\nu\ct-\hat F(s)
\nn\\
&=&c_1\ze_H\at 2s-1,\nu+\hat n\ct+c_3\ze_H\at 2s-3,\nu+\hat n\ct\,,
\label{BBB7} \eeq $\hat n$ being the number of terms not satisfying
the condition $\hat\la_n>|\al|$. In order to proceed, we have to
compute the quantities $\hat b_k$ and $\hat a_k$ for $k=0,1,2$. To
this aim, we note that Hurwitz functions have only a simple pole at
1 and, more precisely, \beq
\ze_H(s+1,\nu)=\frac{1}{s}-\psi(\nu)+O(s)\,, \label{BBB8} \eeq
$\psi(s)$ being the logarithmic derivative of Euler's gamma
function. After a straightforward computation, we get \beq \hat
b_0=\hat\ze(0)=c_1\,\ze_H\at-1,\nu\ct+c_3\,\ze_H\at-3,\nu\ct\,,
  \hs \hat b_1=\frac{c_1}2\,,
   \hs \hat b_2=\frac{c_3}2\,,
\label{bk}\eeq
\beq
\hat a_0&=&\hat\ze'(0)-\ga\hat\ze(0)
\nn\\&=&c_1\,\aq2\ze'_H\at-1,\nu\ct-\ga\ze_H\at-1,\nu\ct\cq
+c_3\,\aq2\ze'_H\at-3,\nu\ct-\ga\ze_H\at-3,\nu\ct\cq\,,
\\
\hat a_1&=&-c_1\,\aq\psi\at\nu\ct+\frac{\ga}{2}\cq
     +c_3\,\ze_H\at-1,\nu\ct\,,
\\
\hat a_2&=&c_1\,\ze_H\at3,\nu\ct
-c_3\,\aq\psi\at\nu\ct+\frac{\ga-1}{2}\cq\,.
\label{ak}\eeq
Using (\ref{ze1}) we obtain
\beq
\ze'_\al(0|\ell^2L)&=&
\at F_\al(0)+\sum_{k=0}^2\,\frac{b_k\al^k}{k!}\ct
\ln\frac{\ell^2R_0}{12}
\nn\\&&\hs\hs
+\sum_{k=0}^2\,\frac{(a_k+\ga b_k)\al^k}{k!}
+F_\al'(0)+\sum_{k=3}^\ii\,\frac{\al^k\hat G(k)}{k}\,.
\label{BBB9}
\eeq

Now we have to consider separately the operators $L_0,L_1,L_2$ we
are dealing with and explicitly compute $b_k$, $a_k$, and $\hat
G(k)$ using (\ref{hlan}), (\ref{bk})-(\ref{ak}), and
(\ref{L0PP})-(\ref{L2PP}).

\subsection*{The scalar case}

The eigenvalues of $L_0$ are of the form \beq
\la_n=\frac{R_0}{12}\aq\at n+\frac32\ct^2-\al\cq\,,\hs
\al=\frac{9}{4}+q\,,\hs\hs n=0,1,2... \label{BBB10} \eeq This case
is model dependent since the parameter $q$ explicitly depends on the
choice of the Lagrangian $f(R)$. Then one could have zero modes and
also negative eigenvalues, but we take them into account by the
functions $F_\al$ and $\hat F$, both of which will in general appear
in the final result. For $k\geq3$, we have \beq \hat
G(k)=-\frac{1}{12}\,\ze_H\at 2k-1,\frac32+\hat n\ct+
\frac13\,\ze_H\at 2k-3,\frac32+\hat n\ct\,, \label{BBB11} \eeq where
$\hat\la_n>|\al|$ per $n>\hat n$. Then \beq
Q_0(\al)\equiv\ze'_\al(0|\ell^2L_0) &=& \aq\,
N_0-\frac{17}{2880}-\frac{\al}{24}+\frac{\al^2}{12}
\cq\,\ln\frac{\ell^2R_0}{12} \nn\\&&\hs +\frac13\aq
3F_\al'(0)+4\ze'H(-3,3/2)-\ze'_H(-1,3/2)\cq \nn\\&&\hs\hs -\aq
72\hat F(1)+11-6\psi(3/2)\cq\frac{\al}{72} \nn\\&&\hs\hs\hs -\aq
12\,\hat F(2)+4\psi(3/2)+7\ze_R(3)-10\cq\frac{\al^2}{24}
\nn\\&&\hs\hs\hs\hs +\sum_{k=3}^\ii\frac{\al^k\hat G(k)}k\,,
\label{Q0}\eeq $\ze_R(s)$ being the Riemann zeta function. One of
the three scalar Laplacian-like operators appearing in the one-loop
effective action (\ref{Gamma}) does not depend on the model since
for such case $\al=\al_0=33/4$. Then $\hat\la_0$ and $\hat\la_1$ are
smaller than $\al$ ($\hat n=2$) and so \beq
F_\al(s)=(-6)^{-s}+5\,(-2)^{-s}\,,\hs\hs \hat
F(s)=\at\frac{9}{4}\ct^{-s}+5\,\at\frac{25}{4}\ct^{-s}\,,
\label{BBB12} \eeq \beq \hat G(k)= -\frac{1}{12}\,\ze_H\at
2k-1,\frac72\ct+ \frac13\,\ze_H\at 2k-3,\frac72\ct\,. \label{BBB13}
\eeq From these equations it follows \beq Q_0(33/4)\sim
-18.32-6\pi\,i
  +\frac{479}{90}\,\ln\frac{\ell^2R_0}{12}\,.
\label{BBB14}
\eeq
We see that there is an imaginary part since there are
negative eigenvalues.

\subsection*{The vector case}

The eigenvalues of $L_1$ are of the form \beq
\la_n=\frac{R_0^2}{12}\aq\at n+\frac52\ct^2-\al\cq\,,\hs
\al=\frac{13}{4}+q\,,\hs\hs n=0,1,2... \label{BBB15} \eeq For the
vector case, $q=3$ is a pure number and so $\al=25/4=\hat\la_0$.
Thus there is a zero-mode with multiplicity equal to 10 ($N_0=10$)
and this has to be excluded in the evaluation of zeta function. As a
consequence, we have $F_\al(s)=0$, $\hat F(s)=10\,\al^{-s}$,
$b_0=\hat b_0-10$, $a_0=\hat a_0+10\ga$, $a_1=\hat a_1-10\al^{-1}$,
$a_2=\hat a_2-10\,\al^{-2}$, and for $k\geq3$ \beq \hat G(k)=
-\frac{9}{4}\,\ze_H\at 2k-1,\frac72\ct+ \ze_H\at 2k-3,\frac72\ct\,.
\label{BBB16} \eeq Finally, \beq Q_1(25/4)\equiv
\ze'_\al(0|\ell^2L_1)&=&
  -\frac{191}{30}\,\ln\frac{\ell^2R_0}{12}
+\frac{22215}{64}+4\ze'_H(-3,5/2)
\nn\\&&\hs
-9\ze'_H(-1,5/2)
-\frac{39375}{128}\,\ze_R(3)-\frac{175}{32}\,\psi(5/2)
\nn\\&&\hs\hs
+\sum_{k=3}^\ii\frac{\al^k\hat G(k)}k
\nn\\ &\sim&
-18.91-\frac{191}{30}\,\ln\frac{\ell^2R_0^2}{12}\,.
\label{Q1}\eeq

\subsection*{The tensor case}

The eigenvalues of $L_2$ are of the form \beq
\la_n=\frac{R_0^2}{12}\aq\at n+\frac72\ct^2-\al\cq\,,\hs\hs
\al=\frac{17}4+q\,,\hs\hs n=0,1,2... \label{BBB16b}\eeq As for the
scalar case, here also zero-modes could appear and/or negative
eigenvalues, depending on the parameter $q$. Then, in general, we
have to introduce the functions $F_\al(s)$ and $\hat F(s)$. For
$k\geq3$, we have \beq \hat G(k)=\hat\ze(k)=
-\frac{125}{12}\,\ze_H\at 2k-1,\frac72\ct+ \frac53\,\ze_H\at
2k-3,\frac72\ct-\hat F(k)\, , \label{BBB17} \eeq and \beq
Q_2(\al)\equiv\ze'_\al(0|\ell^2L_2)&=&
\aq\,N_0+\frac{8383}{576}-\frac{125\al}{24}+\frac{5\al^2}{12}
\cq\,\ln\frac{\ell^2R_0}{12} \nn\\&&\hs +\frac13\aq
3F'_\al(0)+20\ze'_H(-3,7/2)-125\ze'_H(-1,7/2)\cq \nn\\&&\hs\hs -\aq
72\hat F(1)+535-750\psi(7/2)\cq\frac{\al}{72} \nn\\&&\hs -\aq
324\hat F(2)+540\psi(7/2)+23625\ze_R(3)-28486\cq\frac{\al^2}{648}
\nn\\&&\hs\hs +\sum_{k=3}^\ii\frac{\al^k\hat G(k)}k\,.
\label{Q2}\eeq

\end{document}